\documentclass[12pt]{iopart}

\usepackage{graphicx}
\usepackage{dcolumn}
\usepackage{bm}
\usepackage[T1]{fontenc}
\usepackage[latin1]{inputenc}
\usepackage[english]{babel}
\usepackage{color}
\usepackage{pgf}
\usepackage{multimedia}
\usepackage{mathptmx}
\usepackage{mathrsfs}
\usepackage{amsfonts}
\usepackage{iopams}

\begin{document}

\title{Strain behavior and lattice dynamics in Ni$_{50}$Mn$_{35}$In$_{15}$}

\author{C.~Salazar Mej\'{i}a$^1$, A.~K.~Nayak$^1$, J.~A.~Schiemer$^2$, C.~Felser$^1$, M.~Nicklas$^1$, M.~A.~Carpenter$^2$ }
\address{$^1$ Max Planck Institute for Chemical Physics of Solids, 01187 Dresden, Germany}
\address{$^2$ Department of Earth Sciences, University of Cambridge, Downing Street, Cambridge CB2 3EQ, UK}
\ead{Catalina.Salazar@cpfs.mpg.de}

\date{\today}

\begin{abstract}
The lattice dynamics in the polycrystalline shape-memory Heusler alloy Ni$_{50}$Mn$_{35}$In$_{15}$ has been studied by means of resonant ultrasound spectroscopy (RUS). RUS spectra  were collected in a frequency range $100-1200$~kHz between 10 and 350~K. Ni$_{50}$Mn$_{35}$In$_{15}$ exhibits a ferromagnetic transition at 313~K in the austenite and a martensitic transition at 248~K accompanied by a change of the magnetic state. Furthermore it displays a antiferromagnetic to ferromagnetic transition within the martensitic phase. We determined the temperature dependence of the shear modulus and the acoustic attenuation of Ni$_{50}$Mn$_{35}$In$_{15}$ and compared it with magnetization data. Following the structural softening, which accompanies the martensitic transition as a pretransitional phenomenon, a strong stiffening of the lattice is observed at the martensitic magneto-structural transition. Only a weak magnetoelastic coupling is evidenced at the Curie temperatures both in austenite and martensite phase. The large acoustic damping in the martensitic phase compared with the austenitic phase reflects the motion of the twin walls, which freezes out in the low temperature region.
\end{abstract}

\maketitle

\section{INTRODUCTION}
The multifunctional properties of Heusler-type Ni-Mn based magnetic alloys that undergo a martensitic transformation originate from the interaction between structural and magnetic degrees of freedom. Properties as shape memory \cite{Kainuma2006,Planes_JOPM_2009}, magnetic superelasticity \cite{Krenke_PRB_2007}, magnetocaloric \cite{Liu2012,GhorbaniZavareh2014} and barocaloric effect \cite{Manosa2010} have attracted considerable attention due to their potential use in applications. However, the lattice dynamics and its coupling to magnetic properties has been studied only little in these alloys. It has been experimentally shown that in Ni-Mn-\textit{X} alloys with $X={\rm Ga}$, Al and In the transverse TA$_2$ phonon branch shows a dip at a particular wave number which softens upon decreasing the temperature toward the martensitic transformation from the high temperature cubic phase toward a lower-symmetry martensitic phase \cite{Stuhr1997,Moya2006,Moya2009}. A softening has been further observed in the elastic constants \cite{Moya2009,Manosa1997,Moya2006a} and is typically found in bcc-based materials which undergo a martensitic transformation. This softening reflects the dynamical instability of the cubic lattice against the shearing of the \{110\} planes along the $<1\bar{1}0>$ directions. Together, with the premartensitic transition found in some of the Ni-Mn-Ga alloys, the softening is a pretransitional effect of the martensitic transition \cite{Castan2005}. Additional evidence of magnetoelastic coupling has been provided by  an enhancement of the anomalous phonon softening accompanying the ferromagnetic (FM) ordering \cite{Stuhr1997} and by the change in the elastic constants upon application of a magnetic field \cite{Moya2009,Moya2006a,Gonzalez-Comas1999}. More generally, softening (or stiffening) is a consequence of coupling between the driving order parameter(s) for a phase transition and strain, which means that observed variations of elastic constants should provide insights into both strain relaxational behavior and the underlying lattice dynamics of Ni-Mn-based Heusler compounds that undergo a martensitic transformation.

Recently, we showed that Ni$_{50}$Mn$_{35}$In$_{15}$  exhibits an inverse magnetocaloric effect of $-7$~K, in a field change of 6~T, that arises mainly from a change in the entropy due to the structural transition \cite{GhorbaniZavareh2014}. At 4~K the application of a 20~T field, still induces the martensitic transition and, besides the change in magnetic moment, leads to a relative change in length of 0.8~\% \cite{Nayak2014}. This indicates a strong magnetostructural coupling, that might be exploited in applications as actuators or sensors. The objective of the work presented here was to characterize both static and dynamic strain coupling in this material by following the temperature dependence of elastic and anelastic anomalies which accompany the magnetic and structural transitions. We present data for a polycrystalline sample of Ni$_{50}$Mn$_{35}$In$_{15}$  obtained by resonant ultrasound spectroscopy (RUS), which is a convenient method for measuring changes in elastic constants and acoustic attenuation of small samples, with dimensions of between 1 and 5~mm, in a frequency range $0.1-1$~MHz \cite{Maynard1996,Schwarz2000,Migliori1997}. RUS has also been used in a similar manner to investigate the elastic properties of Cu-Al-Ni, Co-Ni-Al and Ni-Mn-Ga alloys \cite{Landa2008,Heczko2013,Perez-Landazabal2007}.

\section{EXPERIMENT}
Polycrystalline samples of Ni$_{50}$Mn$_{35}$In$_{15}$ were obtained by arc-melting stoichiometric amounts of the constituent elements under argon atmosphere. The ingots were remelted several times to assure a high homogeneity. Subsequently they were encapsulated in a quartz ampoule under argon atmosphere and annealed at $800^\circ$C for 2~h and then quenched in ice water. The high quality of the samples was confirmed by powder x-ray diffraction. Magnetization measurements were carried out in a physical property measurement system (Quantum Design).
For RUS a sample was cut in the form of an approximately rectangular parallelepiped with dimensions $2.92\times1.43\times1.51~{\rm mm}^3$ and mass 43.2~mg. Resonance spectra were collected using two different in-house built systems. In the high-temperature instrument, a sample sits lightly between the tips of alumina rods which protrude into a horizontal Netzsch $1600^\circ$C resistance furnace. The piezoelectric transducers are at the other end of the rods, outside the furnace \cite{McKnight2008}. In the low-temperature instrument the sample sits directly between the transducers and is suspended in an atmosphere of a few mbars of helium gas, within a helium flow cryostat \cite{McKnight2007}. Spectra containing 50.000, 65.000 or 130.000 data points were collected in the frequency range $50-1200$~kHz during cycles of cooling and heating in the low temperature instrument and heating followed by cooling in the high temperature instrument. A period of 20 minutes was allowed for thermal equilibration before data collection at each set point. The frequency, $f$, and width at half height, $\Delta f$, of selected resonance peaks in the primary spectra were fit with an asymmetric Lorentzian function. For a polycrystalline sample, the square of the resonance frequency of each peak scales with some combination of the shear and bulk moduli but, since the resonance modes involve predominantly shearing motions, the variation of $f^2$ effectively reflects that of the shear modulus. The inverse mechanical quality factor is taken to be $Q^{-1} =\Delta f/f$, and is a measure of acoustic attenuation.

\section{RESULTS}

\begin{figure}[h!t]
\begin{center}
  \includegraphics[width=0.9\linewidth]{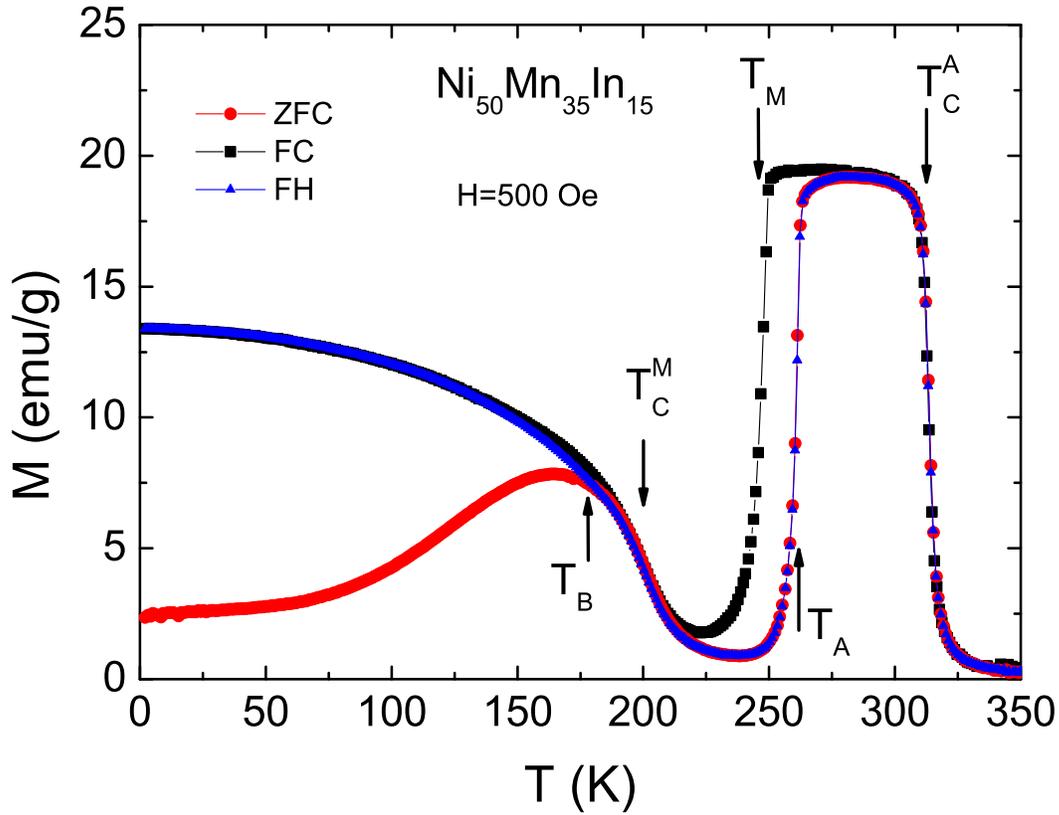}
  \caption{Magnetization curves for Ni$_{50}$Mn$_{35}$In$_{15}$ measured under applied magnetic field of 500~Oe following ZFC (zero field cooled), FC (field cooled) and FH (field heated) protocols. The different transition temperatures are indicated in the figure. See the text for details.}\label{Mag}
  \end{center}
\end{figure}

Figure\ \ref{Mag} shows the magnetization curves for Ni$_{50}$Mn$_{35}$In$_{15}$ measured under an applied magnetic field of 500~Oe following zero-field-cooled (ZFC), field-cooled (FC), and field-heated (FH) protocols. First the sample was cooled down in the absence of field from 350~K down to 2~K. At 2~K the magnetic field was applied and the ZFC curve was measured on heating up to 350~K. Then the FC curve was measured upon cooling and, subsequently, the FH curve was measure on heating. Ni$_{50}$Mn$_{35}$In$_{15}$ exhibits upon cooling a paramagnetic to FM transition in the austenite phase at $T_C^A\approx313$~K, followed by a first-order martensitic magnetostructural transition from a cubic high-temperature phase to a low-temperature monoclinic phase at $T_M\approx248$~K.  Upon heating this transition takes place at $T_A\approx 261$~K. The martensitic transition is accompanied by a change from the ferromagnetic (FM) state to an antiferromagnetic (AFM) state that then orders ferromagnetically at $T_C^M\approx200$~K. The difference in magnetization between the austenite and martensitic phases arises from the changes in the spacing between Mn atoms since the magnetic moments are localized mainly on these and the  exchange interaction strongly depends on the Mn-Mn distance. Hence, any change in the distance caused by a change in the crystallographic configuration can modify the strength of the interactions, leading to different magnetic exchanges in each of the phases \cite{Krenke_PRB_2007}.

Below $T_C^M$, FC and ZFC magnetization curves split at $T_B\approx178$~K. This splitting can be related to the anisotropy of the FM state below $T_C^M$, due to the reduced symmetry of the martensitic phase with respect to the cubic austenitic phase, which leads to a decrease in the number of magnetization easy axes \cite{Krenke2006}. A further explanation is based on the presence of AFM components that pin the FM matrix in different spin configurations depending on whether a cooling-field is present or absent, leading to a magnetically inhomogeneous state \cite{Krenke_NM_2005}.

\begin{figure}[h!t]
\begin{center}
 \includegraphics[width=0.9\linewidth]{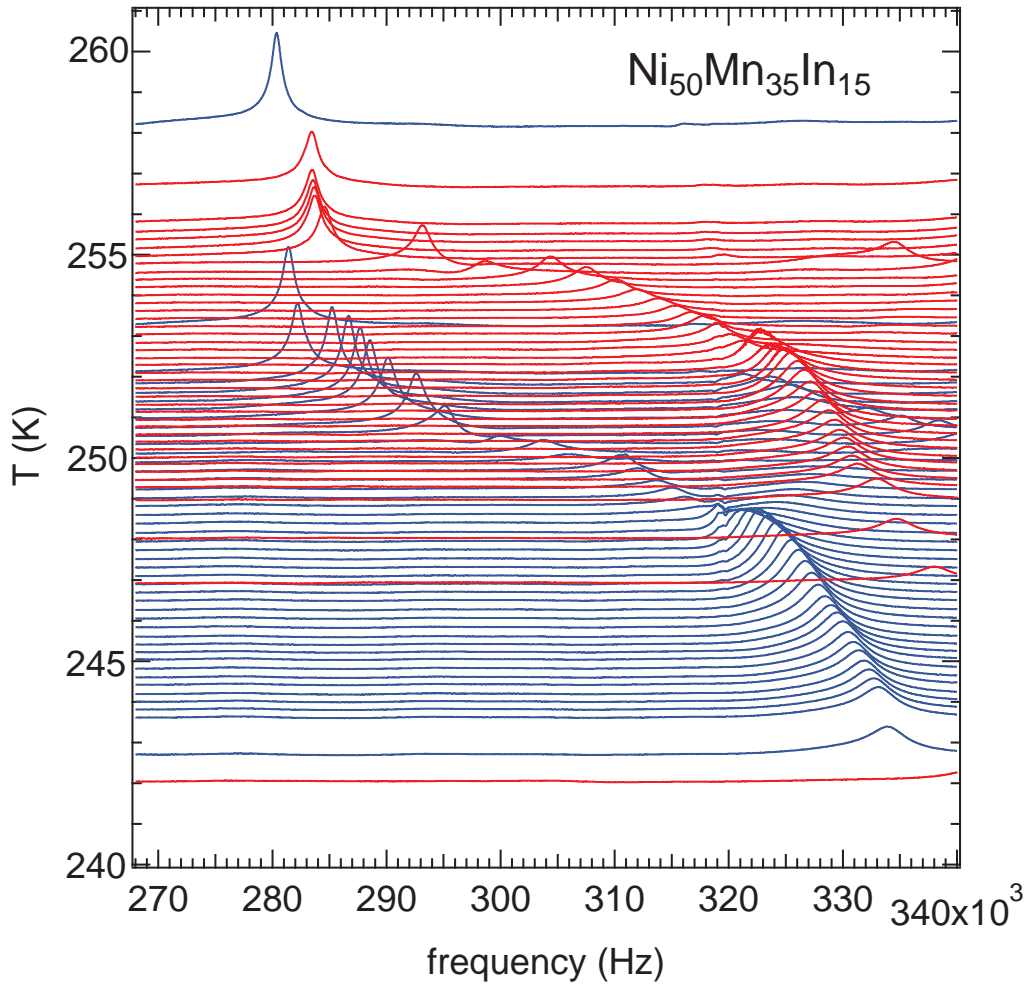}
  \caption{RUS spectra taken at different temperatures through the martensitic transition. The $y$ axis is amplitude in volts, but each spectrum has been offset in proportion to the temperature at which it was collected and the axis is labeled as temperature. Blue traces are spectra collected during cooling and red traces are spectra collected during heating.}\label{spectra}
\end{center}
\end{figure}
\begin{figure}[h!t]
\begin{center}
  \includegraphics[width=0.9\linewidth]{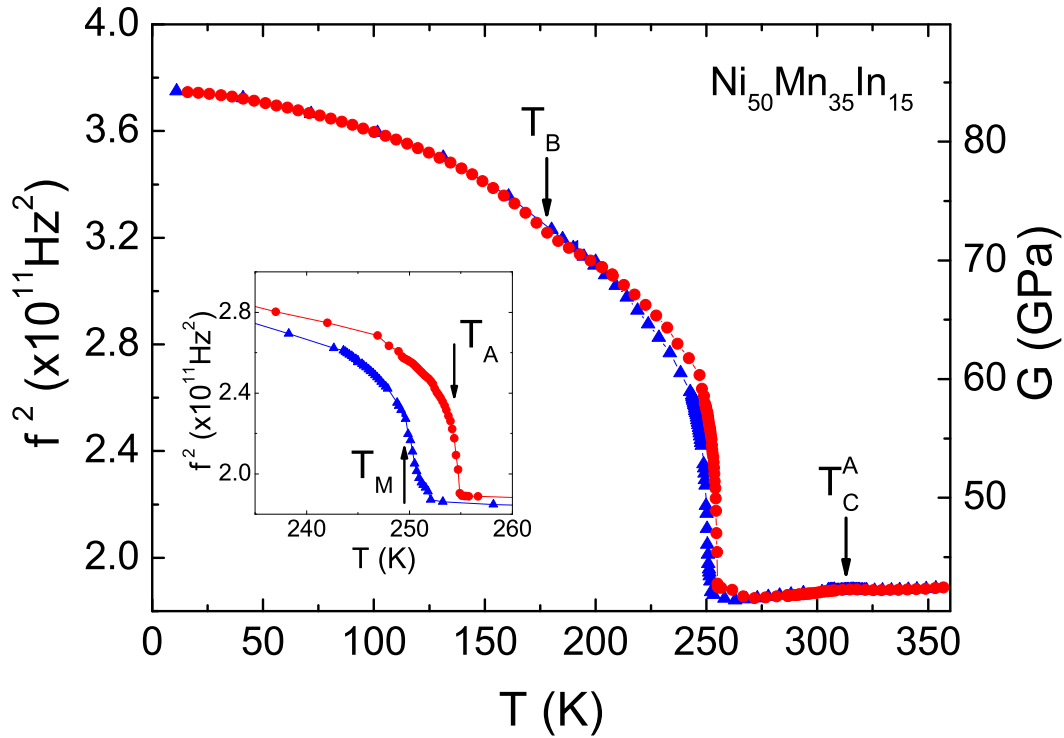}
  \caption{$f^2(T)$ (left axis) obtained from fitting different peaks in the RUS spectra of Ni$_{50}$Mn$_{35}$In$_{15}$. The different $f^2(T)$ curves have been scaled to match the $f^2(T)$ curve for the 0.28~MHz peak. Blue and red color indicates data taken upon cooling and heating, respectively. The right axis shows the variation of the absolute value of the shear modulus obtained by scaling $f^2(T)$ to literature data at room temperature \cite{Moya2009}. The inset magnifies the region around the martensitic transition.}\label{RUS1}
\end{center}
\end{figure}
\begin{figure}[t!]
\begin{center}
  \includegraphics[width=0.9\linewidth]{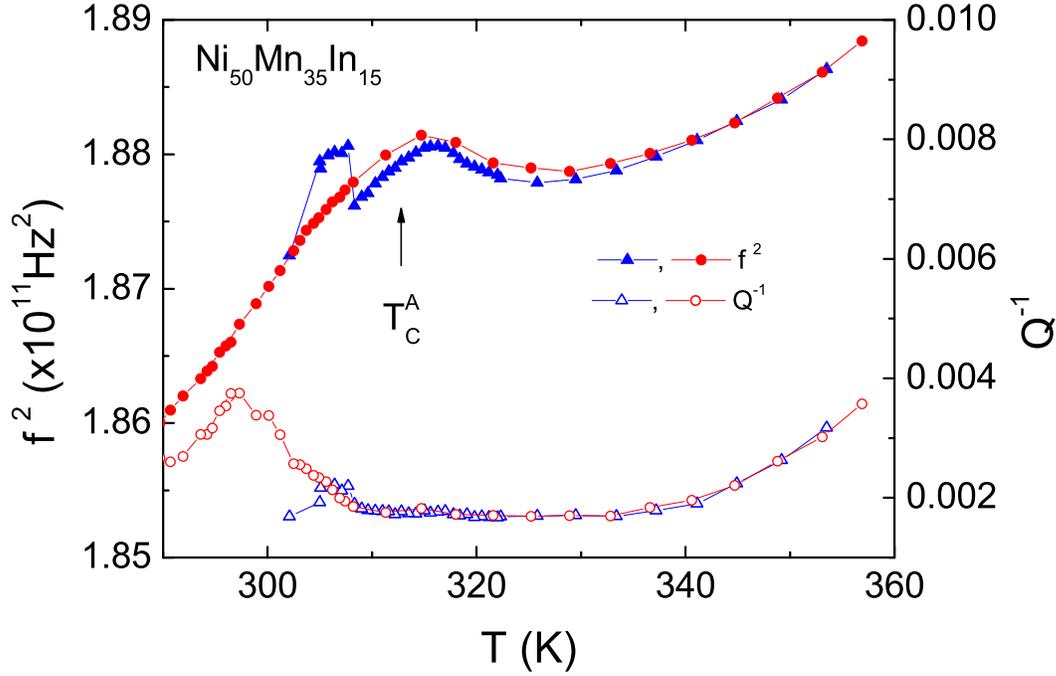}
  \caption{Details of the variation of $f^2$ (closed symbols) and $Q^{-1}$ (open symbols) through the ferromagnetic transition at $T_C^A\approx 313$~K. Blue and red color indicates data taken upon cooling and heating, respectively.}\label{tca}
\end{center}
\end{figure}

Segments of the RUS spectra are shown at narrow temperature intervals through the martensitic transition in Fig.\ \ref{spectra}. They illustrate a change from relatively narrow resonance peaks above the transition point to relatively broad peaks below it, increasing frequency (elastic stiffening) with falling temperature and a hysteresis interval of $5$~K. Variations of $f^2$ for the full temperature range are shown in Fig.\ \ref{RUS1}. They include data from fitting of different peaks which have been combined by scaling to $f\approx0.28$~MHz at room temperature which is the value for the resonance shown in Fig.\ \ref{spectra}. In addition to the large effects seen through the martensitic transition there are clearly also small anomalies at temperatures corresponding to $T_C^A$ and $T_B$. The details of the ferromagnetic transition at $T_C^A$ are presented separately in Fig.\ \ref{tca}, together with the variation of $Q^{-1}$. The shear modulus around the ferromagnetic transition on the austenite displays a bump-like anomaly. A softening of the lattice is observed and it suffers an enhancement in the FM austenite phase.  However, the softening is rather weak and no dip (complete softening) is observed in $f^2(T)$. This behavior is not surprising since no premartensitic transition is present in the Ni-Mn-In Heusler family \cite{Moya2009}.

\begin{figure}[t!]
\begin{center}
  \includegraphics[width=0.9\linewidth]{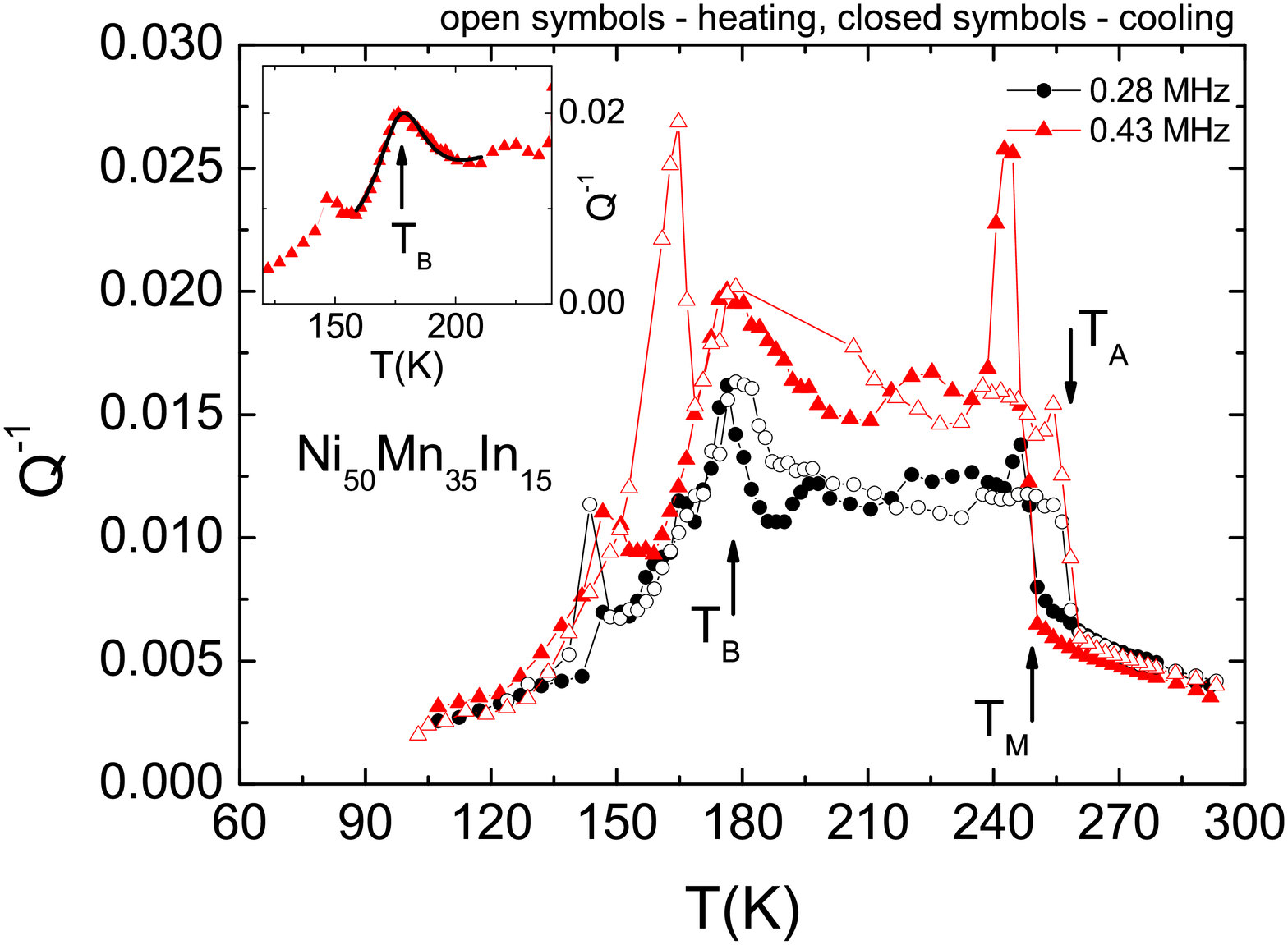}
  \caption{Acoustic loss, expressed as $Q^{-1}$ obtained from fitting resonant peaks with frequencies 0.28~MHz (black) and 0.43~MHz (red), at room temperature, measured on cooling (closed symbols) and heating (open symbols). The arrows indicate the main peaks observed and the transition temperatures determined from magnetization measurements. The inset shows a fit to the Debye peak at $T_B=178$~K from cooling data for the resonance peak with $f\approx0.43$~MHz at room temperature.}\label{invQ}
\end{center}
\end{figure}

Variations of $Q^{-1}$ through the temperature interval $100-295$~K are shown for two resonances, with frequencies $0.28$ and $0.43$~MHz at room temperature, in Fig.\ \ref{invQ}. There is some scatter in the data, but the features reproduced from both peaks and both for heating and cooling are a steep increase with falling temperature at $T_M$ and $T_A$, a plateau of relatively high values down to $200$~K, a peak at $T_B$ and then a decline to values corresponding to those of the high temperature structure by $150$~K.

\section{DISCUSSION}
In combination, the magnetization and RUS data allow a straightforward comparison of the strength of strain coupling associated with each transition. Clearly, the dominant changes in the shear modulus are associated with the large shear strains associated with the martensitic transition. The very small anomaly in $f^2$ at $T_C^A$ implies only weak coupling of the ferromagnetic order parameter with shear strain, and, similarly, for the magnetic ordering at $T_C^M$ there is little or no deflection in the trend of $f^2$ which might indicate any significant magnetoelastic coupling below the second magnetic ordering transition.

The martensitic phase below $T_M/T_A$ in a sample with a close composition (Ni$_{50}$Mn$_{50-x}$In$_x$ at $x = 15.2$) is known to consist of a mixture of 10M and 14M structures \cite{Khovaylo2009}, each of which has substantial shear strains with respect to the parent cubic structure. The magnitudes of these at room temperature can be estimated using lattice parameter data for the 10M structure given by Khovaylo \emph{et al.} \cite{Khovaylo2009}: $a = 4.377$, $b = 5.564$, $c = 21.594$~\AA, $\beta= 91.93^\circ$. If orthogonal reference axes, $X$, $Y$ and $Z$, are chosen as being parallel to the crystallographic $a$, $b$ and $c^\ast$, respectively, the non-zero strain components with respect to a cubic structure with lattice parameter $a_o$ are given by
\begin{eqnarray}
e_1&=\frac{a-a_o}{a_o}\\
e_2&=\frac{b/\sqrt{2}-a_o}{a_o}\\
e_3&\approx\frac{c/5-a_o}{a_o}\\
e_5&\approx \cos \beta
\end{eqnarray}
where an approximation for the value of $a_o$ is
\begin{equation}
a_o\approx \Big( \frac{a\cdot b\cdot c}{5\sqrt{2}}\Big)^{1/3}.
\end{equation}
In symmetry-adapted form, tetragonal and orthorhombic strains are given by
\begin{eqnarray}
e_t&=\frac{1}{\sqrt{3}}(2e_3-e_1-e_2)\\
e_o&=(e_1-e_2).
\end{eqnarray}
On this basis, values of the room temperature shear strains are $e_t = 0.036$, $e_o = 0.091$, $e_5 = -0.034$. As is well known for martensitic transitions these are large in comparison with values in the range $1-3$\% for more typical ferroelastics, and would be expected to give rise to very substantial changes in the shear elastic constants according to the classical model of strain/order parameter coupling described originally by Slonczewski and Thomas \cite{Slonczewski1970}.

There are three order parameters to consider but, as already noted, coupling between the magnetic order parameter(s) and shear strain is evidently weak. This leaves two order parameters driven by changes in electronic structure, the first of which, $Q_1$, would give rise to the symmetry change Fm$\bar{3}$m - I4/mmm and the second, $Q_2$, is responsible for the multiple repeat of the 10M structure. $Q_1$ has the symmetry properties of the irreducible representation $\Gamma^+_3$ and if the complication of incommensurate ordering is ignored $Q_2$ would have symmetry properties corresponding to a point away from the Brillouin zone center along the $k = [\xi,\xi,0]$ line. The essential point is that $Q_1$ would couple bilinearly with a tetragonal strain ($\lambda eQ$) to give pseudoproper ferroelastic behavior and $Q_2$ would have linear-quadratic coupling ($\lambda eQ^2$) to give an improper ferroelastic transition. With regard to the elastic constants, the former would give characteristic softening of ($C_{11} - C_{12}$) as $T$ approaches the martensitic transition point from above and below while the latter would give a step-like softening below the transition temperature \cite{Carpenter1998}. The shear modulus of a cubic crystal comprises of both ($C_{11} - C_{12}$) and $C_{44}$ but there is no real indication in the data for $f^2$ in Fig.\ \ref{RUS1} of significant pseudoproper ferroelastic softening cooling towards the martensitic transition point from the stability field of the cubic phase. It is therefore concluded that the martensitic transition is driven primarily by $Q_2$ and is essentially improper ferroelastic. Rather than there being an abrupt softening below the transition point, however, there is an apparently continuous and large stiffening. From this it is clear that any contributions from relaxations due to $\lambda eQ^2$ coupling must be small, where $e$ could be any of the three shear strains, and the overall elastic properties are determined by the next higher order (biquadratic) terms, $\lambda e^2Q^2$. In this case the amount of elastic stiffening with respect to the cubic parent structure scales with $Q_2^2$. The transition is first order in character, so the apparently continuous variation of the effective shear modulus is most likely to be a consequence of averaging the elastic constants of the two (or more) coexisting structures in the close vicinity of the transition point.

Elastic softening due to coupling of the form ($\lambda eQ^2$) requires that on the time scale of some applied stress there is a relaxation of the strain and, consequently, of the order parameter. It is inevitable that there must be some frequency above which an approach to equilibrium elastic properties is not observed because there is insufficient time for these relaxations to occur. In this case the contributions of higher order coupling of the form $\lambda e^2Q^2$ might still be detected. Such an effect might normally be expected to be small whereas the data in Fig.\ \ref{RUS1} show stiffening of the shear modulus by over 100\%. Similar stiffening has been observed in association with the martensitic transition in single crystal Cu$_{74.08}$Al$_{23.13}$Be$_{2.79}$ both at $0.1$~MHz and $0.25-8$~Hz \cite{Salje2009}, implying that the issue may not be simply a matter of dispersion with respect to frequency, though this is not a group/subgroup transition and the elastic stiffening was attributed to pinning of twin walls by dislocations. Substantial stiffening in an entirely unrelated material, SrAl$_2$O$_4$, also has a pattern that appears to be due to $\lambda e^2Q^2$ coupling \cite{Carpenter2010a}, and is most unlikely to have been affected by dislocations. The hexagonal-monoclinic transition in SrAl$_2$O$_4$ also involves a combination of two order parameters, one of which belongs to a zone center irreducible representation (also pseudoproper ferroelastic), while the other belongs to a zone boundary irreducible representation \cite{Franzen1982,Carpenter1998}. The two non-zero shear strains have values up to about $1$~\% and $6$~\%. The common feature between Ni$_{50}$Mn$_{35}$In$_{15}$  and SrAl$_2$O$_4$ is the presence of two order parameters with different symmetry properties, and we therefore speculate that the strains from these lock together in such a way that strain/order parameter relaxation is suppressed. The outcome could then be a substantial increase in rigidity, i.e. the structures become elastically stiffer rather than softer.

As seen in Fig.\ \ref{tca}, the onset of elastic softening ahead of the martensitic transition is at the ferromagnetic transition temperature, $T_C^A\approx313$~K. The effect is too subtle to be seen in the relatively noisy ultrasonic data of Moya \emph{et al.} \cite{Moya2009}, but a similar pattern of softening occurs in ($C_{11}-C_{12}$) below the equivalent ferromagnetic transition and ahead of the premartensite transition in a single crystal with composition close to stoichiometric Ni$_2$MnGa \cite{Seiner2009,Heczko2012}. Seiner \emph{et al.} reported that the lattice geometry remains cubic in this temperature interval, again signifying that direct magnetoelastic coupling is weak. The form of the elastic anomaly expected from coupling of the form $\lambda em^2$ would be a stepwise softening at $T=T_C^A$, but instead there is continuously increasing softening with respect the trend extrapolated from above $T= T_C^A$, consistent with the coupling coefficient, $\lambda$, being negligibly small. In this case the softening would scale with $m^2$, due to coupling of the form  $\lambda e^2m^2$, as appears to be the case. Softening rather than stiffening implies that the coupling coefficient has opposite sign (and is much smaller) in comparison with stiffening below the martensitic transition temperature.

The variation of $Q^{-1}$ in Fig.\ \ref{invQ} has a pattern which is typical of acoustic loss accompanying a ferroelastic phase transition \cite{Carpenter2015}. The conventional explanation would be of a steep increase at the transition point associated with the appearance of twin walls which are mobile under application of an internal stress. In the austenite phase there is no twinning or any other significant damping mechanism present in  there in contrast to the martensite phase \cite{Aaltio2008}. The plateau of $Q^{-1}$ below this is due to thermally activated mobility of the twin walls in an effectively viscous medium. At some lower temperature the twin walls become pinned by defects in a freezing interval which is readily identifiable by the development of a Debye peak in the loss and an increase in stiffness with respect to shear, as appears to be occur at $T_B$. The loss peak can be fit according to \cite{Carpenter2010}
\begin{equation}\label{debye}
Q^{-1}(T)=Q_{B}^{-1}\Big[\cosh\Big\{\frac{E_a}{r_2(\beta)}\Big(\frac{1}{T}-\frac{1}{T_{B}}\Big)\Big\}\Big]^{-1}
\end{equation}

where $Q^{-1}_B$ is the maximum of the peak at $T_B$, $E_a$ is the activation energy and $r_2(\beta)$ is a width parameter, which arises from any spread in relaxation times for the dissipation processes. The inset of Fig.\ \ref{invQ} shows the results of a fit to $Q^{-1}(T)$ from the peak in the RUS spectra near 0.43~MHz. After subtracting a base line, the fitting gives $Q^{-1}_B = 0.009$, $E_a/r_2(\beta) = 31$~kJ/mol, and $T_B = 177.8$~K. If the loss mechanism is assumed to be due to a single relaxation process ($r_2(\beta) = 1$) related to displacements of domain walls, the Debye peak in $Q^{-1}$ signifies the freezing of the domain-wall movement with a thermal activation energy of about 31~kJ/mol ($0.3$~eV). The peak in $Q^{-1}$ occurs at $\omega \tau= 1$, where $\omega$ is the angular frequency ($= 2\pi f$) and
\begin{equation}
\tau=\tau_o \exp \Big(\frac{E_a}{RT}\Big).
\end{equation}
Using the fit values of $T_B$ and $E_a$, at $f = 5.65 \times 10^5$~Hz then gives an estimate for the inverse attempt frequency as $\tau_o= 2.2\times10^{-16}$~s. Variations of $Q^{-1}$ below the Debye loss peak are noisy but decrease steeply with falling temperature.

The RUS data are consistent with a previous report of pinning of twin walls below $\sim170$~K in Ni-Mn-Ga martensite \cite{Heczko2003}, which is comparable with the freezing behavior found here at $T_B\approx 180$~K. Discussions of twin wall mobility in the 10M phase of Ni-Mn-Ga alloys have focused on the stress felt by two different types of twin walls under the influence of an externally applied magnetic field \cite{OHandle2006,Faran2013,Pramanick2014}. Of these, one has a thermally activated mechanism for sideways displacements in response to the field, and the mechanism is understood to involve nucleation and subsequent migration of ledges along the walls. An activation energy barrier of 0.15-0.30~eV has been reported to give good agreement with observations \cite{OHandle2006}, and the present result is consistent with this. Internal friction measurements at $2$~Hz of a Ni-Mn-Ga alloys with unspecified superstructure types gave activation energy values in the range $0.02-0.04$~eV, however \cite{Gavriljuk2003}. The pinning mechanisms are not understood but the higher value could imply a mechanism involving interaction with impurity atoms \cite{Gavriljuk2003}.

A freezing process described by equation (\ref{debye}) must be accompanied by elastic stiffening according to the Debye equations in the usual way. The relationships required in this context are
\begin{equation}
\tan \delta=\Delta \frac{\omega \tau}{1+\omega^2 \tau^2}
\end{equation}
where $\delta$ is the phase angle, and, in the case of a standard linear solid
\begin{equation}
\Delta=\frac{C_U-C_R}{C_R} \mathrm{\ for\ } (C_U-C_R)\ll C_R
\end{equation}
$C_U$ is the relevant elastic modulus for the unrelaxed state, excluding strain due to movement of the wall, and $C_R$ the modulus of the relaxed state, including the strain due to movement of the wall to its new equilibrium position \cite{Nowick1972}. The relationship between $\tan \delta$ and $Q^{-1}$ is \cite{Lee2000,Lakes2004,Carpenter2010b}
\begin{equation}
\tan \delta\approx \frac{1}{\sqrt{3}}\frac{\Delta f}{f}=\frac{1}{\sqrt{3}}Q^{-1}.
\end{equation}

On this basis, and using $Q^{-1}_B = 0.009$ at $T = T_B$ ($\omega \tau= 1$), the expected change in $f^2$ in Fig.\ \ref{RUS1} is $0.033 \times 10^{11}$~Hz$^2$ which is sufficiently close to the observed change, $0.06\times 10^{11}$~Hz$^2$, to confirm that both anomalies could be due simply to freezing of the twin wall motion. The nature of the pinning mechanism is not known, but it may only be coincidence that it occurs just below the transition from antiferromagnetic to ferromagnetic transition at $T_C^M\approx200$~K. By comparing the $Q^{-1}(T)$ with the ZFC magnetization curve we see that the decrease in the acoustic loss goes along with a decrease in the ZFC magnetization. This behavior might be related with competing AFM and FM interactions which are blocked in a glassy magnetic state \cite{Umetsu2011a}. The strong frequency dependence of the peak below $T_B$ can be taken as a further hint for the glassy character of the strain distribution in the sample. The existence of a strain-glass phase has been proposed in the shape-memory alloys Ti$_{50}$Ni$_{50-x}$Fe$_x$ \cite{Zhang2011} or, recently, in the Heusler material Ni-Co-Mn-Ga \cite{Wang2012}. It has been further predicted for Ni-Co-Mn-\textit{Z} compounds with $Z={\rm In}$, Sn, and Sb \cite{Entel2014}.
Because of the magnetoelastic coupling, supermagnetoelastic behavior (existence of strong FM and AFM interactions in Mn-rich Heusler alloys, which allows elastic softening of the magnetic sublattices, in particular, near the magnetostructural transition), magnetic-cluster-spin glass and strain glass should mutually interact in FM shape memory Heusler alloys. The kinetic arrest phenomenon as a relict of the magnetostructural transformation also supports the glassy behavior and may even lead to strain-glass formation interacting with the magnetic glasses \cite{Entel2014}. The kinetic arrest phenomena has been also observed in Ni$_{50}$Mn$_{35}$In$_{15}$ \cite{SalazarMejia}. Thus, the present finding in RUS data might well be an indication of the presence of a strain-glass phase in in Ni$_{50}$Mn$_{35}$In$_{15}$.

These data for the elastic properties can be compared with data obtained by pulse-echo ultrasonics from a single crystal with composition close to Ni$_{50}$Mn$_{34}$In$_{16}$ by Moya \emph{et al.} \cite{Moya2009}. Although obtained only over a relatively small temperature interval, $200-360$~K, above the martensitic transition there is close agreement with the trend of only slight softening of ($C_{11}-C_{12}$) at $T \rightarrow T_M$. The Voigt-Reuss average, $G_{RV}$, value of the shear modulus obtained from single crystal data at room temperature is 42~GPa. Using this value to calibrate the $f^2$ data in Fig.\ \ref{RUS1}, allows absolute values of the shear modulus to be estimated over the complete temperature range of the RUS data, as depicted on the right axis of Fig.\ \ref{RUS1}, emphasizing the large change in shear stiffness which arises as a consequence of the martensitic transition.

\section{SUMMARY}
The Heusler alloy Ni$_{50}$Mn$_{35}$In$_{15}$  has been studied by means of magnetization measurements and resonant ultrasound spectroscopy which allow a straightforward comparison of the lattice dynamics, the strain behavior and their coupling to the magnetic phase transitions. The dominant changes in the shear modulus are connected with the large shear strains associated with the martensitic transition. A strong stiffening of the lattice is observed accompanied by a marked increase in the acoustic loss. A detailed analysis following the Landau theory reveals that the martensitic transition appears to be driven primarily by the order parameter $Q_2$, which is responsible for the multiple repeat of the 10M structure, and is essentially improper ferroelastic. The stiffening observed could be due to the presence of two order parameters with different symmetry properties where the strains associated with them lock together in such a way that the strain/order parameter relaxation is suppressed. The large damping in the martensite phase in comparison with the austenite phase is related with the presence and mobility of twin boundaries. Our acoustic-loss data indicate a strong decrease of the twin boundaries mobility and hint at a glassy behavior below $T_B\approx178$~K.  Even though, no premartensitic transition takes place in Ni$_{50}$Mn$_{35}$In$_{15}$, pretransitional (premonitory) effects indicated by a weak softening of the shear modulus are present above the martensitic transition. The shear modulus exhibits only small features at the two ferromagnetic transitions reflecting the weak coupling of the ferromagnetic order parameter with shear strain. In this study RUS proved to be a complementary technique to magnetic probes providing insights in the coupling between strain and magnetic degrees of freedom being the basis of the multifunctional properties present in Ni-Mn-based Heusler alloys, which are candidate materials for use in applications as actuators, sensor, or refrigerant materials.

\section*{ACKNOWLEDGMENTS}
This work was in part financially supported by the ERC Advanced Grant (291472) "Idea Heusler". RUS facilities in Cambridge have been supported by grants from the Natural Environment Research Council (NE/B505738/1, NE/017081/1) and the Engineering and Physical Sciences Research Council (EP/I036079/1).

\section*{References}
%

\end{document}